**Title:**

**Ultra-Hot plasma of magnetic monopoles as fifth phase of matter:**

**Bonding dissociation condition of N-S magnetic in a hot plasma medium and superlattices of solids**

By   M. M. Bagheri-Mohagheghi*,

*School of Physics, Damghan University, Damghan, Iran,

P.O. BOX: 41167-36716.

Corresponding Author:

M. M. Bagheri-Mohagheghi

Email: bmohagheghi@du.ac.ir,  ORCID : 0000-0002-8642-7456

**Abstract:**

A magnetic atom has two magnetic poles of N and S, and in normal conditions, separation of magnetic poles cannot be happened, even in very small atomic dimensions. Existence of magnetic dipoles is due to a fundamental property of magnetic materials and belongs to intrinsic spin of electrons. For the electron, N and S magnetic poles are considered, similar to a current loop. The aim of this paper is theoretical calculate dissociation energy of N-S poles a quantum magnetic particle with two approaches of classical and quantum mechanics by providing a harmonic oscillator simple model to estimate dissociation energy of the N-S poles and corresponding breakdown temperature and internal pressure. The results showed that separation of magnetic poles occurs in two states: (a) in an ultra- hot plasma medium with extremely high temperatures, such as in the center of a hot star, and (b) at extremely high pressures, such as between internal plates in complex superlattices of layered solids. It will be shown that breakdown temperature is in order of $\theta = 10^7$ to $10^8$ Kelvin. This temperature is very high and it only happens in an ultra-hot plasma environment as fifth phase of matter. In addition, based on this model, we calculated that the possibility of dissociation of bonds between N and S magnetic poles for solid superlattices occurs at very high pressures between crystal plates. According to these results, the presence of isolated magnetic monopole in superlattices of solids under ultra-high-pressure conditions is possible. Therefore, this model suggests that the conductivity of magnetic monopole carries can be used in the manipulation of nanomaterials for applications in the production of advanced devices such as new generation of superconductors, new spin devices and magnetic-electronics advanced materials with magnetic monopoles and super-dielectrics.

**1-Introduction:**

A magnetic matter has two magnetic poles of N and S, even in very small atomic dimensions. In normal condition existence of magnetic dipoles is due to a fundamental property of magnetic materials and belongs to electron spin. Under normal environmental conditions, it is not possible to separate the two magnetic poles. But recently, a series of advanced experiments at very high pressures have revealed magnetic monopoles. As shown in **Fig.1**, magnetic monopoles act like positive and negative free electric single charge, creating a magnetic field inward or outward around itself [1-3]. Recently, since 2011, there has been a growing interest in study of the magnetic monopoles and conducting international experiments to detect it [3-5].

On the other hand, "*Duality in nature symmetry*" is provided both electric and magnetic field sources i.e., magnetic monopoles and magnetic currents. Magnetic monopoles, while elusive as elementary particles, exist in many materials in the form of quasiparticle elementary excitations. Magnetic monopoles and associated currents were directly measured in experiments, confirming the predicted symmetry between electricity and magnetism. In this field, the conditions for the formation of magnetic monopoles, electrical conduction, thermal conductivity, and their wave and energy equations are discussed [3,4].

One of the important subjects is symmetry in nature, which predicts the existence of iterant magnetic monopoles as single free charges at low temperatures in Maxwell's equations. According to Dirac, the existence of magnetic monopoles with an electric unit charge of Q= (137) e, from $eg = nhc/2$ Dirac quantization condition (DQC), is possible according to quantum equations due to the convergence of nature and the principles of physical phenomenology [6-8]. However, under natural conditions, it is not possible to create a magnetic monopole. Only with very high energy or high mechanical forces or ultra-high pressures it is possible to break the magnetic dipole into two free magnetic monopoles. In addition, by applying external pressures in the Giga-Pascal range, thermodynamic processes changes and new phenomena can be created in complex solid-state structures under critical conditions. For example, by applying a pressure of a few giga-pascals, very good conductivity can be achieved with crystalline helium, as well as for diamonds at high pressures. Induced pressure changes the initial nature and the physical properties of matter [9-12].

In recent years, much research has been done on spin- ice, in which the presence of magnetic monopoles has been proven in some of the superlattice structures of solids such as pyrochlores structures and superconductor materials. This produces a bound pair of north and south poles, which can be fractionalized into two free magnetic monopoles [12-16]. Currently, spin-ice property in the pyrochlore magnetic materials such as $Yb_2Ti_2O_7$, $Er_2Ti_2O_7$ and $Dy_2Sn_2O_7$ have been studied as host strong quantum fluctuations of magnetic dipoles owing to pseudospin (1/2) in magnetic rare-earth elements [17–19].

Newly, a topological model for study of critical phenomena and phase transitions by *Thouless, and et al* (winners of Physics Nobel Prize 2016) are stablished. In this theory, phase transitions of matter are related to the topological defects i.e., vortices or structural holes. In this model, the topologic phase change in critical phenomena such as superconductivity, magnetic ultra-thin films, and superfluid by *topological defects* are shown [20, 21]. Based on this, magnetic monopoles can be attributed to topological holes.

Research in magnetic monopoles has entered many areas of theoretical and experimental physics, including solid state physics, stars physics, elementary particle physics, gravity, and cosmology [22-25].

In this paper, by presenting a simple mechanical and quantum model based on the energy of a simple harmonic oscillator, we try to investigate the breaking conditions of the magnetic dipole junction and obtain its corresponding temperature and internal pressure in solids. This temperature is very high and it only happens in an ultra-hot plasma environment as fifth phase of matter. On the other, we will see that these conditions can be also created under high mechanical pressures in quantum structures and superlattices of layered solids, which can provide many applications in the future.

## 2. Calculation of dissociation energy of N-S poles

### 2-1. Theoretical model with classical approach

Consider a magnetic particle with N-S magnetic poles. These poles may be modeled as a simple classic harmonic system including two bodies of mass M connected by a spring of elasticity $\gamma$, and relative distance of x as shown in **Fig 2**. With this approach, it is possible to calculate dissociation (binding) energy of N-S poles similar to classical physics model. The schematic view of N-S magnetic poles of a magnetic particle which convert to two isolated body including N-pole and S-pole is shown in **Fig 2(a)** and **(b).**

Consider a two-body oscillator ($M_N$ and $M_S$) that becomes a single-body oscillator with reduced mass $(1/\mu) = (1/M_N+1/M_S)$. If x is the relative distance change in spring length due to F, considering the Newton's second law, an answer for the differential equation of distance (x) of the harmonic system is reached as following:

$$\mu \frac{d^2x}{dt^2} + \gamma x = 0 \qquad (1),$$

$$\omega = \sqrt{\frac{k}{\mu}} \qquad (2),$$

x(t) = A sin ωt            (3)

For calculating the oscillation frequency (ω) or bonding energy of dipoles, it is necessary to have spring elasticity constant ($\gamma$). The bonding energy of magnetic poles is considered to have a relation similar to Coulombic interaction force ($F_{N-S}$) between two electric charges with a magnitude of $Q_N=Q_S=$ (137e) with up spin and down spin, based on the Dirac's prediction [7,8] for a magnetic monopole according to quantum calculations as shown in equation (4) and **Fig 3**.

$$F_{N-S} = K \frac{Q_N Q_S}{x^2} \quad (4)$$

In this relation, K is columbic constant of environment which equals $1/4\pi\varepsilon$ and x is magnetic monopoles relative distance which is in sub-atomic range in order of electron's effective radius and we are considered as $10^{-12}$ m. **Fig.2** shows this magnetic quantum particle with single magnetic charges of $Q_N$ and $Q_S$ equal to 137e, with ↑ spin up for N pole and ↓ spin down for S pole. On the other hand, from the classical mechanic equation of the simple harmonic oscillator, this Coulomb force is equal to the elastic force of the oscillator with elastic constant ($\gamma$), as brought in (5).

$$F = K \frac{Q_N Q_S}{x^2} = -\gamma x \quad (5)$$

It is possible to calculate elastic constant ($\gamma$) of the spring according to (5) and then total energy of the system can be obtained by equal consideration of potential and kinetic energy, as brought in equations (6) and (7).

$$\gamma = K\frac{Q_N Q_S}{x^3} \qquad (6)$$

The total energy of the magnetic bipolar system according to the classic law of equivalence of energy kinetic energy ($E_K$) and potential energy (U) in kinetic theory of gases is:

$$E = E_k + U = 2U = 2 \times (\frac{1}{2}\gamma x^2) = \gamma x^2 \qquad (7)$$

With replacement of $\gamma$ from equation 6 in energy equation 7, total energy of magnetic bipolar from equation (8) will be obtained.

$$E = K\frac{Q_N Q_S}{x^3} \times x^2 = K\frac{Q_N Q_S}{x} \qquad (8)$$

As an example, for this approximation, the dissociation energy of the magnetic dipole in quantum scale is calculated by replacing approximate value of x in equation (8).

If is dielectric constant ($k$) = $10^3$ for a typical dielectric material, for example lead zirconate titanate (see Table1), K=$\frac{1}{4\pi k \epsilon_0}$ = $10^{-3} \times (9 \times 10^9)$ and $x$ is relative change in spring length, we can calculate total energy with replacing $Q_N$ and $Q_S$ in equation (8) in two state:

(a)- If x is considered as $10^{-12}$ m (as many ten effective electron radius), energy (E) will be estimated equal to $2.7 \times 10^4$ electron volt (eV) as following:

E= $(9 \times 10^6)\frac{(137\,e)(137\,e)}{10^{-12}}$

E= $9 \times (137)^{2} \times (10^{6}) \times (1.6 \times 10^{-19})^2 / 10^{-12}$

= $432437.76 \times (10^{-20})$ = $4.32 \times 10^{-15}$ J, or ≈ $2.7 \times 10^4$ eV. (9)

As obtained in equation (9), this energy value is equivalent to thermal energy at a temperature of 2×10⁸ Kelvin, existing in core of hot stars:

$$E = \frac{3}{2} k_B \theta_B, \quad k_B \text{ (Boltzmann constant)} = 1.38 \times 10^{-23} \text{ J/K} \quad \longrightarrow \quad \theta = \frac{E}{\frac{3}{2} K_B} = 2 \times 10^8 \text{ K}. \quad (10)$$

(b)- If x is considered as $10^{-9}$ m (1nm) for example for magnetic quantum dot, the corresponding temperature for equivalent thermal energy will be in scale of $10^5$ Kelvin. This result is in conflict with the reality that the corresponding temperature is in scale of $10^8$ Kelvin.

These results show that the equivalent energy is highly dependent to mean distance between magnetic dipole.

**2-2. Theoretical model with quantum approach**

Considering the system as a quantum harmonic oscillator with quantum approach is also possible. For a quantum oscillator with reduced mass of (μ) we have:

$$E = \left(n + \frac{1}{2}\right)\hbar\omega, \quad \omega = \sqrt{\frac{\gamma}{\mu}} \quad (11)$$

Minimum energy (base state) of the oscillator for n = 0 can be supposed as:

$$E = \left(\frac{1}{2}\right)\hbar\omega \quad (12)$$

By replacing $\gamma$ from (6) in (11) for ω and with considering reduced mass μ = M/2 and M = 137$m_e$, we obtain:

$$E = \left(\frac{1}{2}\right)\hbar\omega = \left(\frac{1}{2}\right)\hbar\sqrt{\frac{\gamma}{\mu}} = \left(\frac{1}{2}\right)\hbar\sqrt{\frac{K\frac{Q_N Q_S}{x^3}}{\mu}} \quad (13).$$

$$E = \left(\frac{1}{2}\right) \hbar \sqrt{\frac{2K\frac{Q_N Q_S}{x^3}}{M}} \quad (14)$$

According to previous discussions, to obtain an approximate estimation for energy of quantum oscillator, we have $Q_N = Q_S = (137)$ e and mean distance $x = 10^{-12}$ m. Now we can calculate equivalent temperature as following:

$$E = \left(\frac{1}{2}\right) \hbar \sqrt{\frac{2K\frac{Q_N Q_S}{x^3}}{M}}$$

$$\hbar = 1.054 \times 10^{-34} J.s$$

$$E = \left(\frac{1}{2}\right) \times 1.054 \times 10^{-34} \times \sqrt{\frac{2 \times (9 \times 10^6)\frac{(137\ e)(137\ e)}{10^{-36}}}{M}}$$

$$E = \left(\frac{1}{2}\right) \times 1.054 \times 10^{-34} \times \sqrt{\frac{2 \times (9 \times 10^6)\frac{(137)^2(1.6 \times 10^{-19})^2}{10^{-36}}}{137 \times 9.11 \times 10^{-31}\ (kg)}}$$

$$E = \left(\frac{1}{2}\right) \times 1.054 \times 10^{-34} \times \sqrt{\frac{2 \times (9 \times 10^6)\frac{(137)\ (1.6 \times 10^{-19})^2}{10^{-36}}}{9.11 \times 10^{-31}\ (kg)}}$$

$$E = 4.38 \times 10^{-16} J = 2.74 \times 10^3\ eV. \quad (15)$$

$E = \frac{3}{2} k_B \theta_B$, $k_B$ (Boltzmann constant) = $1.38 \times 10^{-23}$ J/K,

$$\theta = \frac{4.38 \times 10^{-16}\ J}{\frac{3}{2} \times 1.38 \times 10^{-23}} = 2.1 \times 10^7$$

$$\theta = 2.1 \times 10^7\ K, \quad (16)$$

In quantum harmonic oscillator model, breakdown temperature ($\theta$) is in order of $10^7$ which is in acceptable agreement with mechanical approach results. Of course, separation between magnetic poles is not possible under normal conditions, and only exists in hot stars, which are a hot plasma environment and matter is at ultra-hot plasma phase. Here, we define the fifth state of matter, in which matter becomes so hot that after ionization, as the temperature rises, the magnetic poles separate. In this phase change, in the hot plasma environment of matter, the particles are magnetic monopoles. **Fig.4** shows the phase change from the solid state to the hot plasma phase (the fifth state of matter) as the temperature increases.

## 3. Calculation of dissociation energy of N-S poles in solids with layered structures.

There is another possibility for separation between magnetic poles in superlattice of layered nanostructured solids, which create extremely high pressures between the plates.

However, from thermodynamic equations is the relationship between internal pressure (P) and energy density (E) in solids as follows [26]:

$$P = -(\partial E/\partial V)_N \quad (17)$$

$$P = 2/3 \times \frac{E}{V}. \quad (18)$$

In this relation E and V are energy density and volume of solid, respectively. In this case, we can use Equations 17 and 18 to obtain the approximate value of the pressure between the planes in solids superlattice, which is an example of energy value at equation 15.

If $V = a^3$ and a is lattice constant $= 5 \times 10^{-10}$ m (5 Å) for a typical solid lattice, therefore for energy $E \approx 10^{-16} J$, pressure (P) is in order of $\approx 10^{12}$ pascal, and of course, this amount of pressure is possible in between of plates in super lattices. **Fig.5** show formation conditions of magnetic dipoles under ultra-high pressure between atomic layers of superlattice of solids: (a) before bond breaking and (b) for formation of free magnetic monopoles. Indeed, N-S dipoles between atomic layers under ultra-internal high pressure convert to two single charges after bond breaking and a super-current have produced due to mobility of free magnetic monopoles.

## 4. Conclusions

In this paper, we study the breaking conditions of a magnetic dipole. Herein, it was found that in one simple classical model of harmonic oscillator can be isolated magnetic monopole provided by applying very high energy or temperature and pressure. In the quantum view, we reach the same results in a similar way with a good approximation. The results showed that separation of magnetic poles as fifth phase of matter occurs in two states: (a) in a very hot plasma environment with extremely high temperatures, such as in the center of a hot star, and (b) at extremely high pressures, such as between internal plates in complex superlattices of layered solids.

The most important results are presented as follows:

(a) The simple harmonics oscillator model in two approaches of classical and quantum mechanics is quite successful in calculating the breaking energy of a magnetic dipole particle.

(b) This model suggests that it is possible to create magnetic monopoles in hot plasma media. The breakdown temperature ($\theta$) is in order of $10^7$ and these results have acceptable agreement in both classical and quantum approaches.

(c) In addition, based on this model, we calculated that the possibility of dissociation of bonds between N and S magnetic poles for solid superlattices occurs at very high pressures between crystal plates.

(d) For a typical solid superlattice, for energy $E \approx 10^{-16} J$, pressure is in order of $P \approx 10^{12}$ pascal, and this amount of pressure is possible in between of plates in super lattices such as perovskite and pyrochlore.

This model can be applied for construction feasibility of new spin devices and magnetic-electronics advanced materials with magnetic monopole carriers using the layered superlattice solids.

-------------------------------------------------------------------------------------------------------

------------------------------------------------------------------------------------------------

Table 1. Relative permittivity of some materials at room temperature [26]

| Material | Dielectric constant ($\epsilon_r$) |
|---|---|
| **Pyrex (Glass)** | 4.7 (3.7–10) |
| **Diamond** | 5.5–10 |
| **Graphite** | 10–15 |
| **Silicon** | 11.68 |
| **Glycerol** | 42.5 (at 25 °C) |
| **Titanium dioxide** | 86–173 |
| **Strontium titanate** | 310 |
| **Barium strontium titanate** | 500 |
| **Barium titanate** | 1200–10,000 (20–120 °C) |
| **Lead zirconate titanate** | 500–6000 |
| **Conjugated polymers** | 1.8–6 up to 100,000 |
| **Calcium copper titanate** | >250,000 |

**Figure captions:**

**Fig 1. (a)** Magnetic dipoles and magnetic field of (B) around it and **(b)** Magnetic monopoles and magnetic field of (B) around it and (c) magnetic moments of an electron in the opposite directions [1].

**Fig 2. (a)** a single magnetic particle with N-S poles and **(b)** a two-mass oscillator with $M=M_N=M_P$.

**Fig 3.** System containing: pole N with 137 electrons of (mean) spin up and pole S with 137 electrons of down (mean) spin down.

**Fig .4:** The phase change of matter from solid to ultra-hot plasma with increasing temperature.

**Fig. 5(a):** Magnetic dipoles under ultra-high pressure between atomic layers before bond breaking and **(b)** Free magnetic monopoles between atomic layers after bond breaking and mobility of free monopoles.

(a)

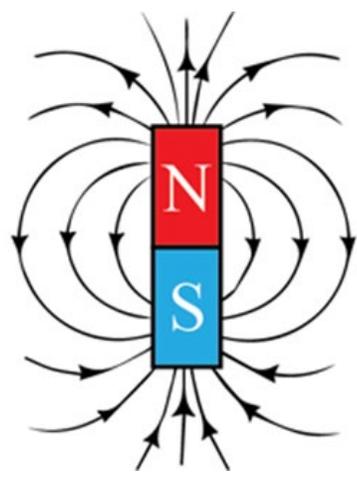

(b)

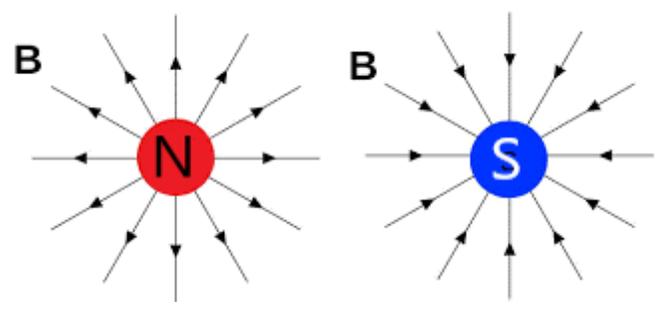

(c)

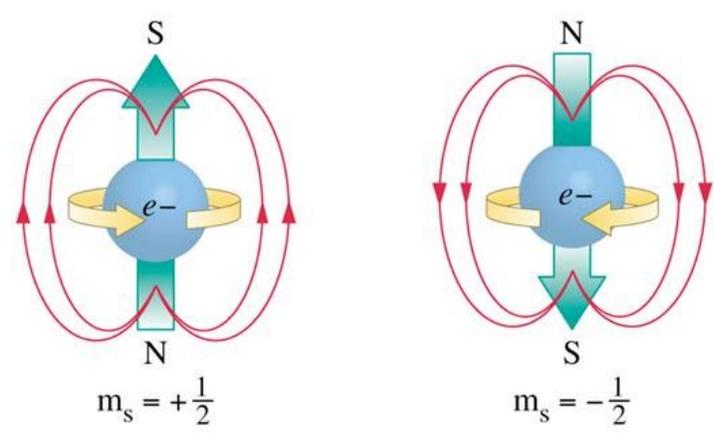

**Fig 1**: (a) Magnetic dipoles and magnetic field of (B) around it and (b) Magnetic monopoles with isolated separated monopoles of S and N and magnetic field of (B) around it and (c) magnetic moments of an electron in the opposite spin directions [1].

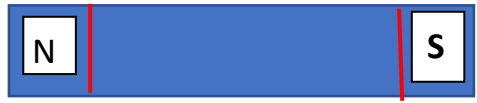

**Fig 2 (a)**: a single magnetic particle with N-S poles.

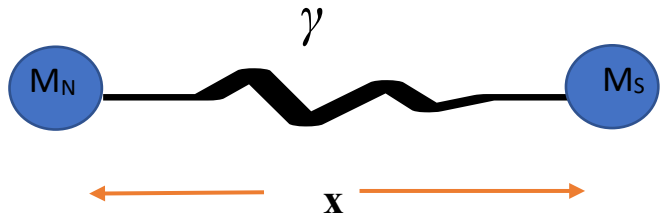

**Fig 2(b)**: A two-mass oscillator with $M=M_N=M_S = 137\, m_e$ with relative distance of x.

-----------------------------------------------------------------------------------------------------------------

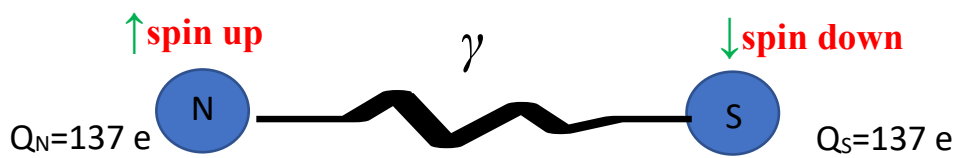

**Fig 3**: System containing: pole N with 137 electrons of (mean) spin up ↑ and pole S with 137 electrons of down (mean) spin down ↓.

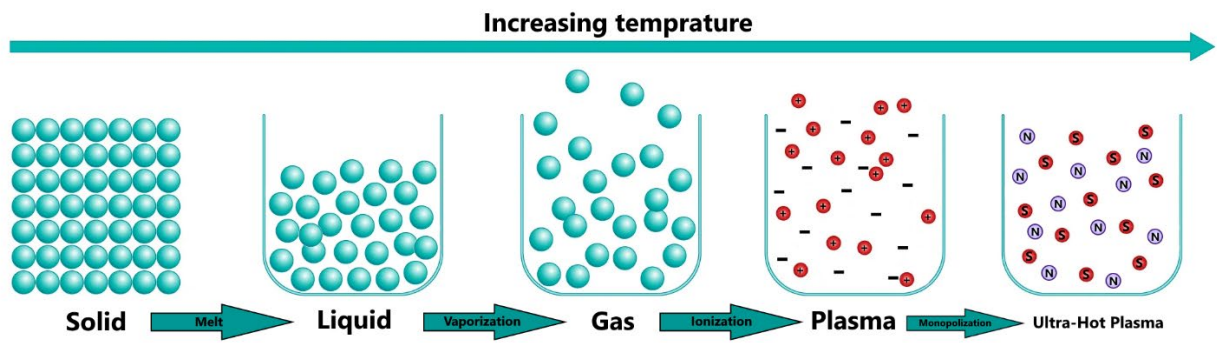

**Fig .4:** The phase change of matter from solid to ultra-hot plasma with increasing temperature.

**(a): Before dissociation of N-S bonding**

**Atomic Layers under ultra-high pressure**

$P = F/A$

**Bonded Magnetic Dipoles**

$P = F/A$

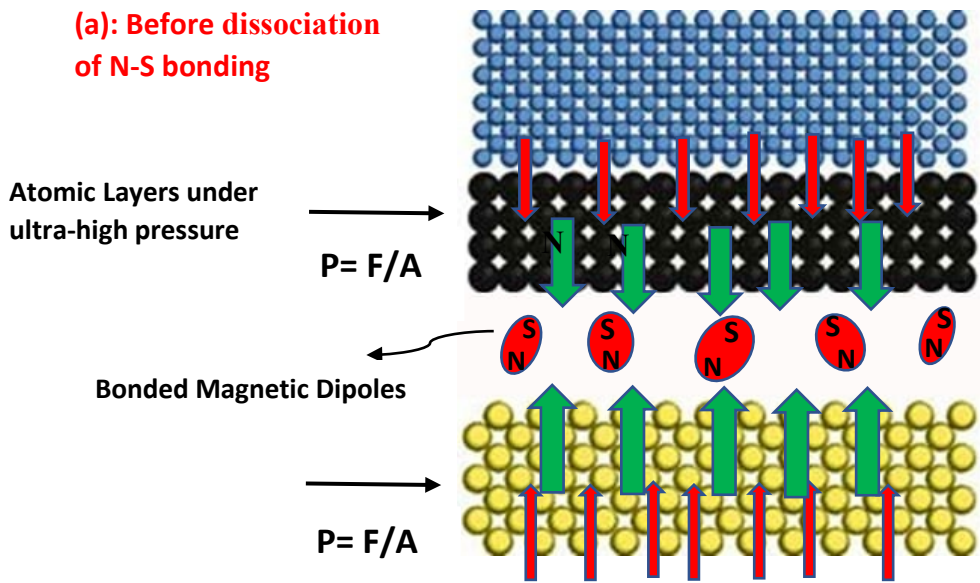

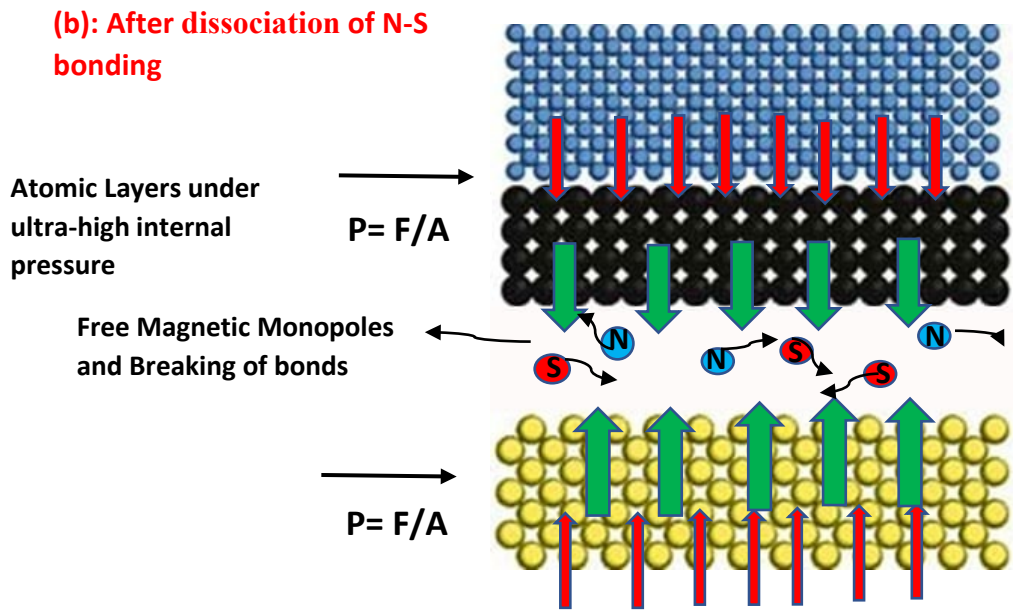

**Fig. 5:** (a) Magnetic dipoles under ultra-high pressure between atomic layers before bond breaking and (b) Free magnetic monopoles between atomic layers after bond breaking and mobility of free monopoles.